\documentclass[prd,preprint,showpacs,superscriptaddress,showkeys,floatfix,nofootinbib]{revtex4}
\pdfoutput=1
\usepackage{graphicx}
\usepackage{dcolumn}
\usepackage{bm}
\usepackage{color}
\usepackage{url}
\usepackage{amsmath,amssymb,graphicx}
\usepackage{amsmath}
\usepackage{amssymb}
\usepackage{mathrsfs}
\usepackage{accents}
\usepackage{epsfig}
\usepackage[lofdepth,lotdepth,caption=false]{subfig}
\usepackage[11pt]{moresize}
\usepackage{mathtools}
\usepackage{mathrsfs}
\newlength\figureheight 
\newlength\figurewidth

\begin{document}

\title{Probing Large Extra Dimensions With IceCube}
\author{Arman Esmaili}
\email{arman@ipm.ir}
\affiliation{Institute of Convergence Fundamental Studies \& School of Liberal Arts,
Seoul National University of Science and Technology, Seoul 139-743, Korea}
\affiliation{Instituto de F\'isica Gleb Wataghin - UNICAMP, 13083-859, Campinas, SP, Brazil}
\author{O. L. G. Peres}
\email{orlando@ifi.unicamp.br}
\affiliation{Instituto de F\'isica Gleb Wataghin - UNICAMP, 13083-859, Campinas, SP, Brazil}
\affiliation{Abdus Salam International Centre for Theoretical Physics, ICTP, I-34010, Trieste, Italy}
\author{Zahra Tabrizi}
\email{tabrizi.physics@ipm.ir}
\affiliation{School of Particles and Accelerators, Institute for Research in Fundamental Sciences (IPM), P.O.Box 19395-1795, Tehran, Iran}
\affiliation{Instituto de F\'isica Gleb Wataghin - UNICAMP, 13083-859, Campinas, SP, Brazil}

\date{\today} 

\begin{abstract}
In models with Large Extra Dimensions the smallness of neutrino masses can be naturally explained by introducing gauge singlet fermions which propagate in the bulk. The Kaluza-Klein modes of these fermions appear as towers of sterile neutrino states on the brane. We study the phenomenological consequences of this picture for the high energy atmospheric neutrinos. For this purpose we construct a detailed equivalence between a model with large extra dimensions and a $(3+n)$ scenario consisting of three active and $n$ extra sterile neutrino states, which provides a clear intuitive understanding of Kaluza-Klein modes. Finally, we analyze the collected data of high energy atmospheric neutrinos by IceCube experiment and obtain bounds on the radius of extra dimensions.  
\end{abstract}

\maketitle

\section{Introduction\label{sec:int}}

The large extra dimension (LED) model has been introduced and motivated as a solution to hierarchy problem~\cite{ArkaniHamed:1998rs,ArkaniHamed:1998nn,Antoniadis:1998ig}, that is the huge difference between the Planck and weak scales. The basic idea is the confinement of the Standard Model (SM) particles on a brane embedded in the bulk space (that is the space including the extra dimensions), except the graviton which can propagate into the bulk~\cite{Rubakov:1983bb}. In this scenario the \textit{fundamental} Planck scale in the bulk is suppressed down to the weak scale by the volume of extra dimension space and so there is no hierarchy problem anymore. 

In the same scenario, the same idea has been proposed to explain the smallness of neutrino masses~\cite{Dienes:1998sb,ArkaniHamed:1998vp}. In fact, the mechanism of confinement of SM particles on the brane relies on the gauge flux conservation which necessitate that just singlets under the SM gauge symmetry can propagate into the bulk. Thus, in principle, in addition to the graviton, the hypothesized right handed neutrinos can also live in the bulk and consequently the volume suppression explains the small neutrino masses. However, Kaluza-Klein (KK) expansion of the right handed neutrinos after the compactification of extra dimensions manifest towers of sterile neutrinos from the brane point of view which can dramatically affect the oscillation phenomenology of active neutrinos and have been studied extensively in the literature\footnote{For possible signatures of bulk KK modes at colliders and also their impact on the lepton number violating processes see~\cite{Cao:2004tu}. A review of the collider signatures is given in~\cite{Gingrich:2009az}.}~\cite{Dvali:1999cn,Machado:2011jt}. Although the majority of studies derive more and more stringent upper bound on the radius of extra dimensions, still, interestingly, with the current upper limit on the size of extra dimensions, the first KK mode sterile neutrino can have a mass $\mathcal{O}(1)$~eV which is in the ballpark of what is required for the interpretation of the recently observed anomalies in short baseline neutrino experiments and LSND/MiniBooNE experiments~\cite{Kopp:2013vaa}. For instance, in this line, in~\cite{Machado:2011kt} it is proposed that the reactor and gallium anomalies can be interpreted within the LED model.   

In this paper we study an independent probe of the LED model by the use of high energy atmospheric neutrinos. During the past few years, the completed IceCube detector at the south pole collected a high statistics sample of atmospheric neutrino data with energies $>10$~GeV, which actually play the role of \textit{background} for astrophysical/cosmic neutrino searches that IceCube is intended to do. However, these background data provide a unique opportunity to probe new physics scenarios with unprecedented precision, and has been already used to probe sterile neutrinos~\cite{Esmaili:2012nz,Esmaili:2013vza,Esmaili:2013cja,Razzaque:2011ab}, violation of equivalence principle~\cite{Esmaili:2014ota}, non-standard neutrino interactions~\cite{Esmaili:2013fva} and matter density profile of the Earth~\cite{Agarwalla:2012uj}. In this paper we study the signature of LED model in high energy atmospheric neutrinos and, by analyzing the data sets IC-40~\cite{Abbasi:2010ie} and IC-79~\cite{Aartsen:2013jza}, we show that it is possible to constrain the radius of extra dimension to $<4\times10^{-5}$~cm (at $2\sigma$ C.L.). Also, we estimate the sensitivity of IceCube to the LED model after taking into account the energy information of collected data and show that the favored region of parameter space by reactor and gallium anomalies~\cite{Machado:2011kt} can be excluded by IceCube data.    

From the brane point of view the KK modes of LED model resemble a series of sterile neutrino states with increasing masses. The Earth's matter density induce resonant conversion of active neutrinos to these sterile states, which the rate of conversion depends on the energy and zenith angle ($\theta_z$) of atmospheric neutrinos. Phenomenologically, these signatures are similar to the signatures of $(3+n)$ scenarios consisting of three active neutrinos and $n$ sterile states with mixing pattern determined by various mixing angles. We elaborate on this similarity and establish a detailed equivalence between them. 

The paper is organized as follows: In Section~\ref{sec:formalism} we explain the formalism of LED model and the matter effects of the Earth on the KK modes. In Section~\ref{sec:osc} we calculate the flavor oscillation probabilities of high energy atmospheric neutrinos in the LED model. Then in Section~\ref{sec:3+n} we establish the equivalence between the LED and $(3+n)$ models. Section~\ref{sec:icecube} is devoted to the analysis of the data of IceCube. We summarize our conclusions in Section~\ref{sec:conc}.

\section{Matter effects on neutrino propagation in Large Extra Dimensions model\label{sec:formalism}}

In this section we study the propagation of neutrinos in matter in the LED model. Our aim is to investigate the Earth's matter effects on the propagation of high energy atmospheric neutrinos in the presence of Kaluza-Klein modes. The collected data of high energy atmospheric neutrinos by the IceCube detector provides a unique opportunity to search for these effects and so to probe the LED model.

The number of LEDs should be $D\geq 2$, where the $D=1$ case is excluded by the observed $1/r^2$ behavior of the gravitational force at the scale of solar system. The factor suppressing the 4-dimensional Planck scale down to $\sim$ TeV scale is the volume of the $D$-dimensional space, where for the case that LEDs are compactified on tori with radii $R_j$ ($j=1,\ldots,D$), it is given by $(2\pi)^D R_1\cdots R_D$. It should be noticed that all the radii $R_j$ are not necessarily equal, and in fact, assuming an {\it asymmetrical} compactification in the $D$-dimensional space, a hierarchical pattern of $R_j$ elevates the existing bounds on the size of LED radii from supernovae cooling and the cosmological considerations~\cite{Kaloper:2000jb}. A $(4+D)$-dimensional space with hierarchical radii of compactification in the $D$-dimensional space of LED effectively is equivalent to a $5$-dimensional bulk space with the LED radius given by the largest $R_j$ which will be denoted by $R_{\rm ED}$ hereafter. The LED scenario explains the smallness of active neutrino masses through the volume suppression of the Yukawa couplings between the Higgs field $h$, the active left-handed neutrinos $\nu_{iL}$ and the 5-dimensional fermions $\Psi_i$ (singlet under the SM gauge group) where $i=1,2,3$ correspond to the number of active flavors. The action of interaction between the active neutrinos and $\Psi_i$ fields is given by $S= \sum_{i=1}^3 S_i$, where~\cite{Dienes:1998sb}
\begin{eqnarray}\label{eq:action}
S_i=\int d^4x~dy~i\overline{\Psi}_i\Gamma^A\partial_A\Psi_{i}+\int d^4x~\left[ i\bar\nu_{iL}\gamma^{\mu}\partial_{\mu}\nu_{iL}+\lambda_{ij}h\bar\nu_{iL}\psi_{jR}(x,y=0)\right]+\rm{h.c.}
\end{eqnarray}
In this equation $\Gamma_A ~(A=0,\ldots,4)$ are the Dirac matrices and $(\psi_{iL},\psi_{iR})$ are the Weyl components of the fermion $\Psi_i$ living in the 5-dimensional space $(x^\mu,y)$. The first term of Eq.~(\ref{eq:action}) is the kinetic term of $\psi_{iL}$ and $\psi_{iR}$ fields and the first term in bracket is the kinetic term of active neutrino fields $\nu_{iL}$. The last term is the Yukawa term with the coupling constant $\lambda_{ij}$ (with dimension $({\rm mass})^{-D/2}$) which gives the interaction of $\Psi_i$ fields in the bulk with the active neutrinos living on the brane $y=0$ (we are assuming compactification on a $Z_2$ orbifold where $\psi_{iL}$ and $\psi_{iR}$ are odd and even under its $Z_2$ action, respectively; and so $\psi_{iL}(x,y=0)$ vanishes.) The mixings of active neutrinos are parametrized with the PMNS matrix $U$ through
\begin{equation}
\nu_{\alpha L} = \sum_{i=1}^3 U_{\alpha i} \nu_{iL}~,
\end{equation}
where $U\to U^\ast$ for antineutrinos. Without loss of generality, the Yukawa coupling matrix $\lambda_{ij}$ can be diagonalized by the above field redefinition and a corresponding redefinition of the bulk fields. After electroweak symmetry breaking and expansion of the $\psi_{iR}$ and $\psi_{iL}$ fields in terms of the Kaluza-Klein modes, the mass terms of action in Eq.~(\ref{eq:action}) take the following form~\cite{Dienes:1998sb,Dvali:1999cn}
 \begin{equation}\label{eq:massterms}
\sum_{n=-\infty}^{\infty}m_i^D \bar{\nu}_{iL} \psi_{iR}^{(n)}+ \sum_{n=1}^{\infty} \frac{n}{R_{\rm ED}} \left( \overline{\psi_{iL}^{(n)}} \psi_{iR}^{(n)} - \overline{\psi_{iL}^{(-n)}} \psi_{iR}^{(-n)}\right) + \rm{h.c.},
\end{equation}
where $\psi_{iR}^{(n)}$ and $\psi_{iL}^{(n)}$ are the $n^{\rm th}$ KK mode of the bulk fermions $\psi_{iR}$ and $\psi_{iL}$, respectively. The $m_i^D$ are the three mass parameters that form the diagonal Dirac mass matrix $m_{\rm diag}^D$ in this basis, which in turn results from the diagonalization of the matrix $v\lambda_{ij}/\sqrt{V_D}$, where $v$ is the vacuum expectation value of Higgs field and $V_D$ is the volume of compactified space. Let us define the following basis of fields:
\begin{eqnarray}\label{eq:newbasis}
\nu^{(0)}_{iR} & = &  \psi^{(0)}_{iR},\nonumber\\
\nu^{(n)}_{iR} & = &  \frac{\psi^{(n)}_{iR}+\psi^{(-n)}_{iR}}{\sqrt 2},~~~~n=1,\ldots,\infty,\nonumber\\
\nu^{(n)}_{iL} & = &  \frac{\psi^{(n)}_{iL} -\psi^{(-n)}_{iL}}{\sqrt 2},~~~~n=1,\ldots,\infty,
\end{eqnarray}
and the combinations orthogonal to $\nu^{(n)}_{iR}$ and $\nu^{(n)}_{iL}$ which since they decouple from the system we ignore them. In this basis the mass terms in Eq.~(\ref{eq:massterms}) can be written as $\overline{L}_i M_i R_i$, where $L_i^T=\left(\nu_{iL},\nu^{(n)}_{iL}\right)$, $R_i^T=\left(\nu^{(0)}_{iR},\nu^{(n)}_{iR}\right)$ and
\begin{equation}\label{eq:totmassmatrix}
M_i=\lim_{n\to\infty}
\begin{pmatrix}
m_i^D & \sqrt{2}m_i^D & \sqrt{2}m_i^D & \sqrt{2}m_i^D & \ldots & \sqrt{2}m_i^D\\
0 & 1/R_{\rm{ED}} & 0 & 0 & \ldots & 0\\
0 & 0 & 2/R_{\rm{ED}} & 0 & \ldots & 0\\
\vdots & \vdots & \vdots & \vdots& \ddots & \vdots\\
0 & 0 & 0 & 0 & \cdots & n/R_{\rm{ED}}
\end{pmatrix}.
\end{equation}
As can be seen the mass matrix $M_i$ is not diagonal and so we will call the basis of $L_i$ and $R_i$ as ``pseudo-mass" basis.

The Schr\"{o}dinger-like evolution equation of the whole physical states, that is the active neutrinos and the KK modes $\nu^{(n)}_{iL}$, including the matter potentials (which in our case are induced by the Earth's matter) can be written in the pseudo-mass basis as ($k=1,2,3$)\\
\begin{equation}\label{eq:evolution}
\left[ i \frac{d}{dr} L_k = \frac{1}{2E_\nu} M_k^\dagger M_k L_k+ \sum_{j=1}^3
\begin{pmatrix}
X_{kj} & 0_{1\times n} \\
0_{n\times 1} & 0_{n\times n}
\end{pmatrix} L_j \right]_{n\to\infty}~,
\end{equation} 
where $X_{kj} = \sum_{\alpha} U_{\alpha k}^\ast U_{\alpha j} V_\alpha$, and
\begin{equation}
V_\alpha = \delta_{e\alpha} V_{\rm CC} + V_{\rm NC} = \sqrt{2} G_F \left( \delta_{e\alpha} n_e  - \frac{n_n}{2} \right)~,
\end{equation}
where $n_e$ and $n_n$ are the electron and neutron number density profiles, respectively. The same evolution equation applies to antineutrinos with the replacement $X_{kj} \to -X_{kj}$.

An immediate interpretation of the set of evolution equations in Eq.~(\ref{eq:evolution}) is that, from the brane ($y=0$) point of view, the KK modes $\nu^{(n)}_{iL}$ (for each $i$, and $n=1,2,\ldots$) constitute a tower of sterile neutrinos which their masses (and also the masses of active states $\nu_{iL}$) can be obtained by the diagonalization of the matrix $M_i^\dagger M_i$. The matrices $M_i^\dagger M_i$ can be diagonalized by changing the basis from pseudo-mass basis $L_i=(\nu_{iL},\nu_{iL}^{(n)})^T$ to the ``true" mass basis $L^\prime_i=(\nu_{iL}^\prime,\nu_{iL}^{\prime(n)})^T$, where $L^\prime_i = S^\dagger_i L_i$ and $S^\dagger_i M_i^\dagger M_i S_i = (M_i^\dagger M_i)_{\rm diag}$. The active flavor neutrino states $\nu_{\alpha L}$ can be expanded in terms of the ``true" mass basis as
\begin{equation}\label{eq:nualpha}
\nu_{\alpha L} = \sum_{i=1}^3 U_{\alpha i} \nu_{iL}  = \sum_{i=1}^3 U_{\alpha i} \sum_{n=0}^\infty S_i^{0n} \nu_{iL}^{\prime(n)}~,
\end{equation}
where $S_i^{0n}$ is the $0n$ element of the matrix $S_i$ and we defined $\nu_{iL}^{\prime(0)}\equiv\nu_{iL}^{\prime}$. The eigenvalues $\left(\lambda_i^{(n)}\right)^2$ of the matrices $R_{\rm ED}^2 M_i^\dagger M_i$ are the roots of the following transcendental equation~\cite{Dienes:1998sb}
\begin{equation}\label{eq:eigenvalue}
\lambda_i-\pi \left(m_i^DR_{\rm ED}\right)^2 \cot(\pi\lambda_i)=0~.
\end{equation}
So the mass\footnote{These are the masses in vacuum. The matter potentials will modify these masses in the usual way.} of each state $\nu^{\prime(n)}_{iL}$ in $L_i^\prime$ is $\lambda_i^{(n)}/R_{\rm ED}$. The matrix elements $S_i^{0n}$ are given by~\cite{Dienes:1998sb}
\begin{equation}\label{eq:s0n}
\left(S^{0n}_i\right)^2=\frac{2}{1+\pi^2 \left(m_i^DR_{\rm ED}\right)^2 + \left(\lambda^{(n)}_i\right)^2 /\left(m_i^DR_{\rm ED}\right)^2}~.
\end{equation}
It can be shown that Eq.~(\ref{eq:eigenvalue}) has infinite number of solutions $\lambda^{(n)}_i$ where $n < \lambda^{(n)}_i < n+0.5$. Thus, the masses of KK modes $\nu^{\prime(n)}_{iL}$ ($n\neq0$) are increasing roughly as $\sim n/R_{\rm ED}$, while the contribution of KK modes to the active flavor states (that is $S_i^{0n}$) decreases by increasing $n$ (it can be shown that $S_i^{0n}\simeq\sqrt{2}m_i^D R_{\rm ED}/n$~\cite{Dvali:1999cn}). The decrease of the active-sterile mixings by the increase of $n$ means that the higher KK modes gradually decouple from the evolution equation in Eq.~(\ref{eq:evolution}), and so for an experimental setup sensitive to a known energy range we need to consider only a finite number of the KK modes. In the following we discuss the number of KK modes that should be considered for the analysis of the IceCube atmospheric neutrino data.      

In the high energy range ($E_\nu \gtrsim 0.1$~TeV) the Earth's matter effects dramatically change the oscillation pattern of atmospheric neutrinos in the LED model. The matter potentials modify the oscillation phases which lead to resonant conversion of the active neutrinos to the KK modes comprising the tower of sterile neutrinos with increasing masses. The resonance condition in the $2\nu$ approximation of $\nu_{iL}^{(0)}-\nu_{iL}^{\prime(n)}$ system with the effective mixing angle denoted by $\vartheta_n$ is
\begin{equation}\label{eq:res}
\frac{\left(\lambda^{(n)}_i\right)^2 - \left(\lambda^{(0)}_i\right)^2}{2E_\nu R_{\rm ED}^2} \cos2\vartheta_n = V_\alpha~.
\end{equation}  
Due to the sign of $V_\alpha$ for the Earth's matter ($V_e>0$, while $V_\mu, V_\tau < 0$), the resonance condition in Eq.~(\ref{eq:res}) can be fulfilled for $\nu_e$, $\bar{\nu}_\mu$ and $\bar{\nu}_\tau$; which means that at energies satisfying the condition in Eq.~(\ref{eq:res}) the $\nu_e$ ($\bar{\nu}_{\mu/\tau}$) converts to the sterile flavor KK mode $\nu_{sL}^{(n)}$ ($\bar{\nu}_{sL}^{(n)}$). The atmospheric neutrino flux at high energies is dominated by $\nu_\mu$ and $\bar{\nu}_\mu$ with the $\nu_e$ and $\bar{\nu}_e$ components suppressed at least by a factor of ~$\sim20$~\cite{Honda:2006qj}. Also, in this paper we analyze the so-called muon-track events in IceCube which originate from the charged current interactions of $\nu_\mu$ and $\bar{\nu}_\mu$ with the nuclei in the detector. Thus, the main signature of the LED model in the high energy atmospheric neutrinos is in the muon-flavor survival probabilities. Before passing, let us mention two points. Firstly, in the LED model for each flavor of the active neutrinos (or equivalently for each mass eigenstate) there is a tower of KK modes. So, just by considering the first mode ($n=1$) three different mass-squared differences can be inserted in Eq.~(\ref{eq:res}), which lead to three different resonance energies. However, for $R_{\rm ED} \lesssim 10^{-4}$~cm the first KK mode masses are large enough (for reasonable values of $m_i^D$) such that all the active-sterile mass-squared differences are almost equal and effectively there is just one mass-squared difference for each $n$. The current upper limit on $R_{\rm ED}$ from oscillation experiments is $\sim10^{-4}$~cm~\cite{Machado:2011jt} and so the three mass-squared differences for each $n$ are degenerate. Secondly, although for the numerical calculations in sections~\ref{sec:osc} and \ref{sec:icecube} we use the exact position-dependent mass density profile of the Earth from the PREM model~\cite{1981PEPI...25..297D}, in the analytical description of the oscillation pattern we assume a constant average density $\bar{\rho} = 5.5~{\rm g~cm}^{-3}$ for the core-crossing atmospheric neutrinos. The resonances described in Eq.~(\ref{eq:res}) are constant density MSW resonances and the variability of matter density is not playing a significant role except for the core crossing trajectories where the castle wall configuration of mantle-core-mantle leads to the parametric resonances~\cite{Liu:1997yb}.

Let us study the series of resonance energies from Eq.~(\ref{eq:res}). By increasing $n$, ${\cos2\vartheta_n \to 1}$ and $\left(\lambda^{(n)}_i\right)^2 \propto n^2$; so for the resonance energy of conversion to the $n^{\rm th}$ KK mode we obtain $E_\nu^{{\rm res},(n)} \propto n^2$. For $\left(\left(\lambda^{(n)}_i\right)^2 - \left(\lambda^{(0)}_i\right)^2\right)/R_{\rm ED}^2 = 1~{\rm eV}^2$ the resonance energy for core crossing trajectories of $\bar{\nu}_\mu$ (that is $\cos\theta_z=-1$) is\footnote{The MSW resonance energy from Eq.~(\ref{eq:res}) is $\sim4$~TeV. However, for trajectories passing through the core of Earth the parametric resonance dominates at $\sim2.5$~TeV~\cite{Liu:1997yb}.} $\sim 2.5$~TeV. Thus, the series of resonance energies for the atmospheric $\bar{\nu}_\mu$ conversion to the KK modes (assuming $\cos\theta_z=-1$) are
\begin{equation}\label{eq:Eres}
E_{\nu}^{{\rm res},(n)} \simeq 10n^2 ~{\rm TeV} \left( \frac{10^{-5}~{\rm cm}}{R_{\rm ED}} \right)^2~.
\end{equation} 
For the neutrinos passing just the mantle ($\cos\theta_z \gtrsim-0.8$) the resonance energies are $\simeq~16n^2 ~{\rm TeV} \left( 10^{-5}~{\rm cm}/R_{\rm ED} \right)^2$. At high energies ($E_\nu\gtrsim 0.1$~TeV) in the standard $3\nu$ framework the muon-flavor survival probability is $P(\bar{\nu}_\mu\to\bar{\nu}_\mu)=1$; while, qualitatively from Eq.~(\ref{eq:Eres}), in the LED model a series of dips exist at energies $E_\nu^{{\rm res},(n)}$ ($n=1,2,\ldots$), which reflect the conversion of $\bar{\nu}_\mu$ to the $n^{\rm th}$ KK sterile states. The infinite number of resonance energies can be truncated at some $n$ for two reasons: 1) by increasing $n$ the resonance energy increases while the flux of atmospheric neutrinos decreases by the increase of energy as $\propto E_\nu^{-2.7}$. So, the statistics at higher KK modes resonance energies are low and IceCube (or in general any neutrino telescope) would not be sensitive to these KK modes. 2) By the increase of $n$ the mixing between the active and the $n^{\rm th}$ KK mode states decreases ($\sin\vartheta_n\simeq\sqrt{2}m_i^DR_{\rm ED}/n$) which leads to less intense active to sterile conversion. So the depth of resonance dips decrease by the increase of energy and for the large values of $n$ it is beyond the sensitivity reach of the detector. 

In this paper we analyze the atmospheric neutrino data collected during two phases of IceCube construction IC-40~\cite{Abbasi:2010ie} and IC-79~\cite{Aartsen:2013jza} (the numbers mean that at the period of data collection 40 and 79 strings bearing DOMs were deployed, out of the final 86 strings). The energy range of IC-40 and IC-79 data sets are $(0.1-400)$~TeV and $(0.1-10)$~TeV respectively\footnote{The IC-79 data set consists of two high energy and low energy subsets~\cite{Aartsen:2013jza}. For our analysis the high energy subset is relevant which its energy range is $(0.1-10)$~TeV.}. Taking 100~TeV as the energy where above it the statistics are too low, from Eqs.~(\ref{eq:res}) and (\ref{eq:Eres}) the resonance energies are within the energy range of IC-40 and IC-79 for $n\lesssim3~(R_{\rm ED}/10^{-5}~{\rm cm})$. By inserting the current upper limit $R_{\rm ED}\lesssim10^{-4}$~cm it means that at least $\sim30$ KK modes should be taken into account in the calculation of oscillation probabilities. On the other hand, IceCube is sensitive\footnote{This is the sensitivity of IceCube from the analysis of zenith distribution of muon-track events. Adding the energy information improves the sensitivity by a factor of few for resonance energies $\lesssim10$~TeV~\cite{Esmaili:2013vza}. We elaborate more on this in section~\ref{sec:icecube}.} to the active-sterile mixing angles $\sin^22\vartheta_n\gtrsim0.1$~\cite{Esmaili:2012nz}. From Eq.~(\ref{eq:s0n}) this sensitivity can be translated to (assuming $m_i^D R_{\rm ED} \ll 1$)
\begin{equation}\label{eq:nlimit}
n \lesssim 4.5 \left( \frac{R_{\rm ED}}{10^{-5}~{\rm cm}} \right) \left( \frac{\max\left[m_i^D,\sqrt{\Delta m_{\rm atm}^2}\right]}{{\rm eV}} \right)~.
\end{equation}       
The ``max" function in the above relation comes from the fact that if $m_1^D\to 0$, although the mixing between $\nu_{1L}^{(0)}$ and $\nu_{1L}^{(1)}$ vanishes, but the mixing between $\nu_{3L}^{(0)}$ and $\nu_{3L}^{(1)}$ is still sizable because $\lambda_3^{(0)}=\sqrt{\left(\lambda_1^{(0)}\right)^2+R_{\rm ED}^2\Delta m_{\rm atm}^2}$ is not zero. Plugging the current bounds on $m_i^D$ and $R_{\rm ED}$ from~\cite{Machado:2011jt} into Eq.~(\ref{eq:nlimit}) we obtain $n\lesssim3$. Thus, practically very few KK modes contribute substantially to the oscillation pattern of the atmospheric neutrinos. In the above discussion we assumed that all the sensitivity of IceCube to the sterile neutrinos originate from the resonance region; while the interference terms in lower energies are also important and so a few more KK modes should be taken into account. As a conservative assumption, in the numerical calculations of the next section we consider $n=5$ KK modes in the evolution equations.

\section{Numerical calculation of the oscillation probabilities\label{sec:osc}}

The oscillation probabilities of active neutrinos can be found by solving the set of evolution equations in Eq.~(\ref{eq:evolution}). As we mentioned before, for the high energy atmospheric neutrinos which is our interest in this paper, the relevant channel is the survival probability $P(\bar{\nu}_\mu\to\bar{\nu}_\mu)$, or more generally the oscillation probabilities of $\nu_\mu\to\nu_\alpha$ and $\bar{\nu}_\mu\to\bar{\nu}_\alpha$.  

As we discussed and justified in section~\ref{sec:formalism}, we consider $n=5$ KK modes in our numerical calculations. The initial conditions for the calculation of $\bar{\nu}_\mu$ oscillation probabilities in the pseudo-mass basis of Eq.~(\ref{eq:evolution}) are $L_i^j=\delta^{j}_{0} U_{\mu i}^\ast$, where $L_i^j$ is the $j^{\rm th}$ component of $L_i$ and elements of the PMNS matrix $U$ are fixed to their best-fit values~\cite{GonzalezGarcia:2012sz}. The values of $m_i^D$ depend on the mass hierarchy of active neutrinos. For normal hierarchy (NH) $m_2^D= \sqrt{(m_1^{D})^2+\Delta m_{\rm sol}^2}$ and $m_3^D= \sqrt{(m_1^D)^2+\Delta m_{\rm atm}^2}$; and so $m_1^D$ and $R_{\rm ED}$ are the free parameters of the model. For inverted hierarchy (IH) $m_1^D\simeq m_2^D\simeq\sqrt{(m_3^D)^2+\Delta m_{\rm atm}^2}$; and so $m_3^D$ and $R_{\rm ED}$ are the free parameters\footnote{A technical note: To be precise, these relations should be applied to the eigenvalues $\lambda_i^{(0)}$; for example, for inverted hierarchy we would write $\lambda_1^{(0)} \simeq \lambda_2^{(0)} \simeq \sqrt{\left(\lambda_3^{(0)}\right)^2+R_{\rm ED}^2 \Delta m_{\rm atm}^2}$. Then, by knowing the values of $\lambda_1^{(0)}$ and $\lambda_2^{(0)}$ we can calculate $m_1^D$ and $m_2^D$ from Eq.~(\ref{eq:eigenvalue}), which can be used to calculate $\lambda_i^{(n)}$ by the same equation. This procedure have been discussed in detail in~\cite{BastoGonzalez:2012me}. However, in the region of parameter space where we are interested in (and also taking into account the current bounds), it can be shown that applying the mass relations to $m_i^D$ lead to the same results and we can ignore this technical point.}. However, in the high energy range ($E_\nu\gtrsim 0.1$~TeV), since the oscillations driven by $\Delta m_{\rm atm}^2$ and $\Delta m_{\rm sol}^2$ are suppressed and the first KK mode is much heavier than the active neutrino states, the oscillation pattern is the same for both NH and IH and so we show the oscillation probabilities just for NH. Finally, in our numerical calculation, for the matter potential $X_{kj}$ in Eq.~(\ref{eq:evolution}) we used the PREM model~\cite{1981PEPI...25..297D}.

\begin{figure}[t!]
\centering
\subfloat[$P(\bar{\nu}_\mu\to\bar{\nu}_\mu)$]{
\includegraphics[width=0.5\textwidth]{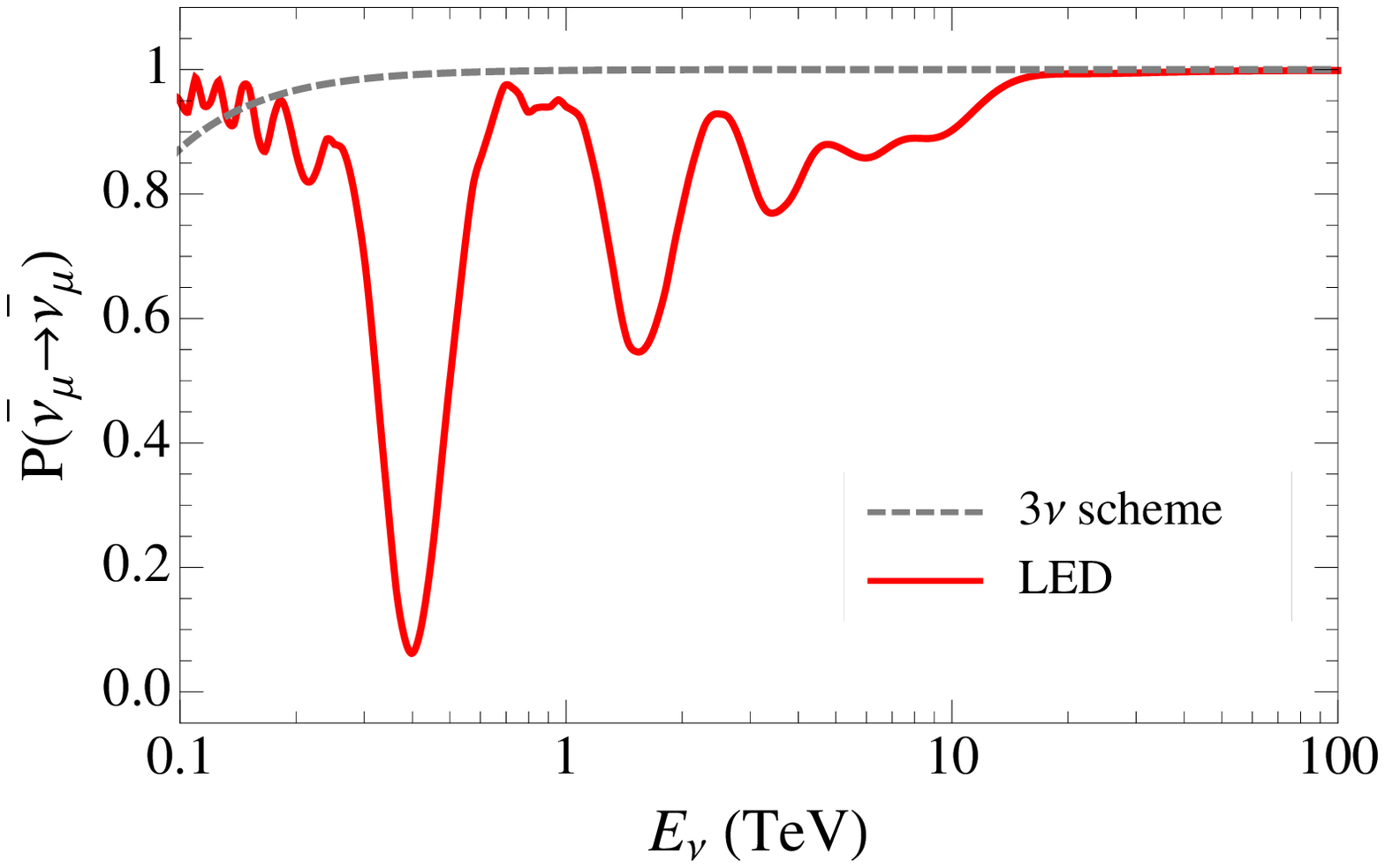}
\label{fig:LEDprob-mubar}
}
\subfloat[$P(\nu_\mu\to\nu_\mu)$]{
\includegraphics[width=0.5\textwidth]{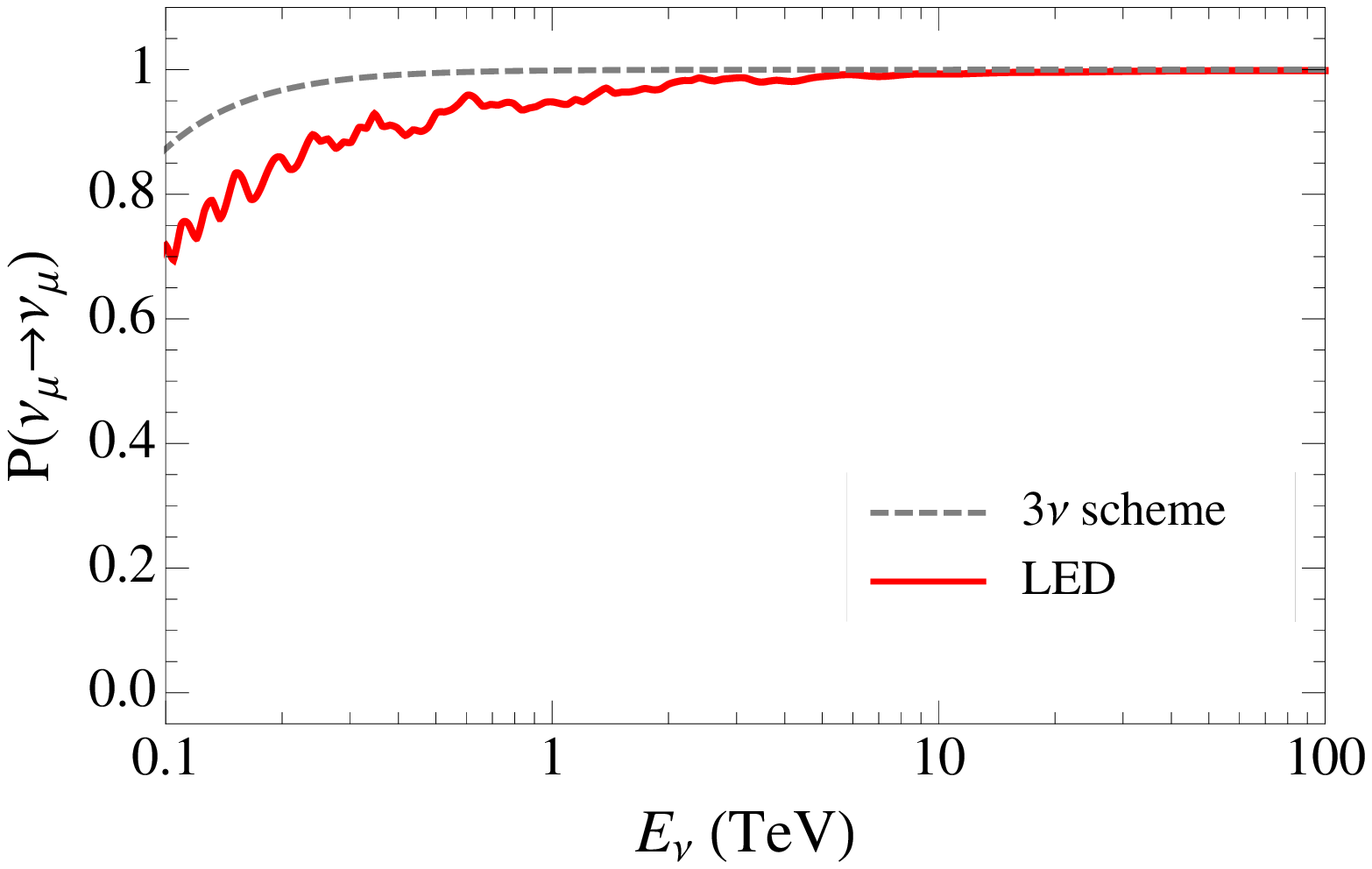}
\label{fig:LEDprob-mu}
}
\quad
\subfloat[$P(\bar{\nu}_\mu\to\bar{\nu}_\tau)$]{
\includegraphics[width=0.5\textwidth]{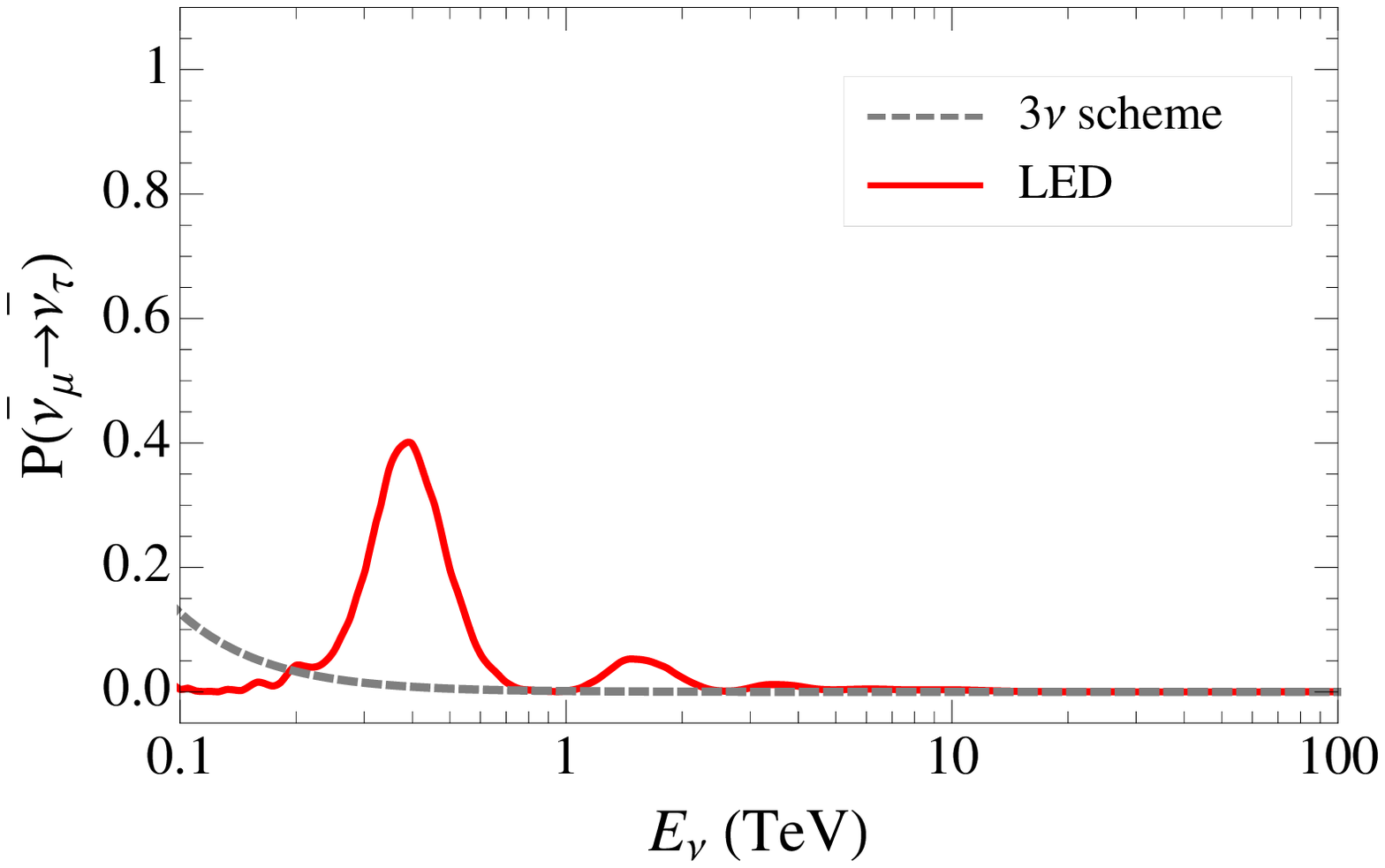}
\label{fig:LEDprob-taubar}
}
\subfloat[$P(\nu_\mu\to\nu_\tau)$]{
\includegraphics[width=0.5\textwidth]{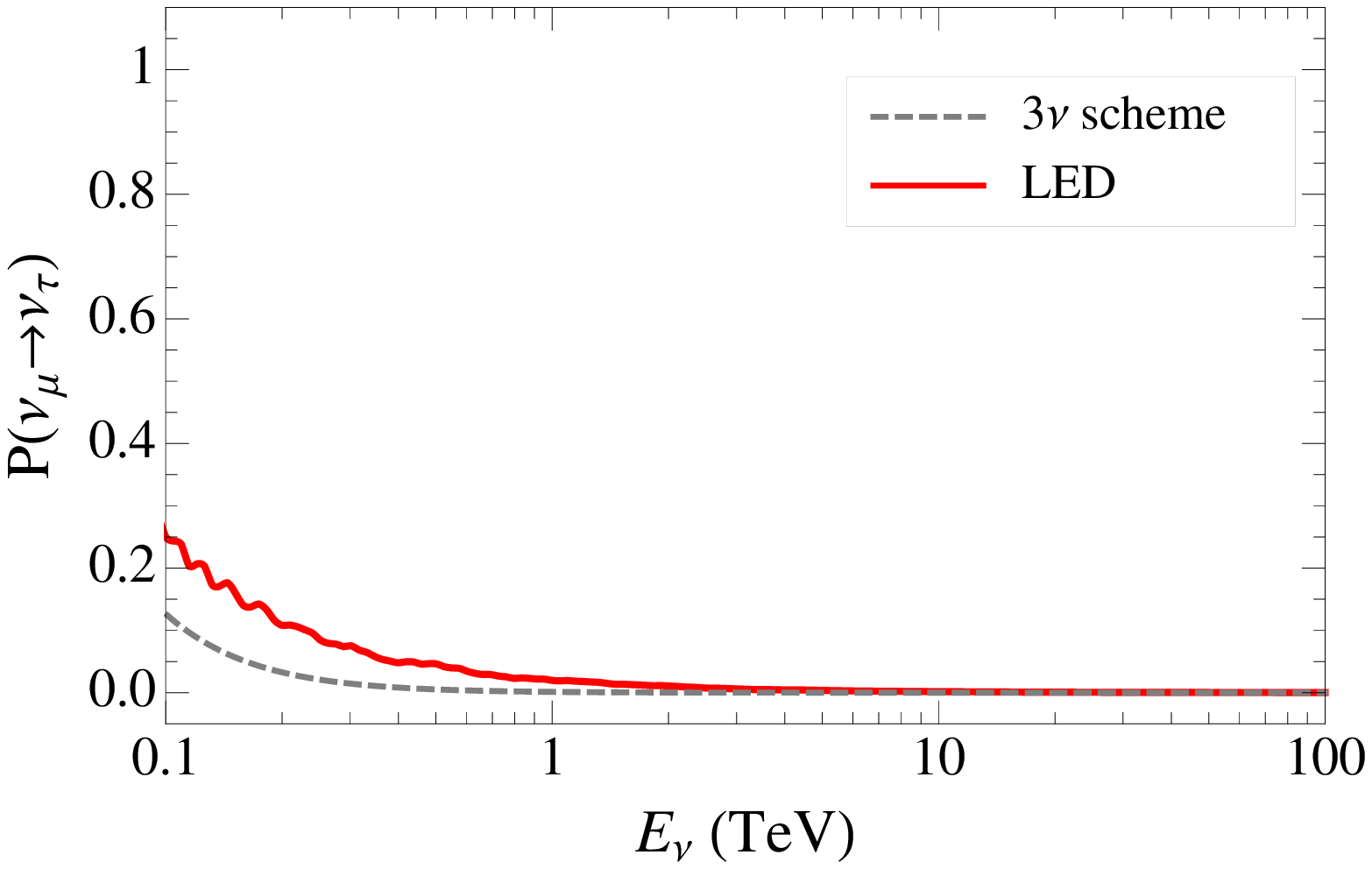}
\label{fig:LEDprob-tau}
}
\caption{\label{fig:LEDprob}The oscillation probabilities as function of neutrino energy $E_\nu$ for $\cos\theta_z=-1$. In all the panels $m^D_1=0.01~\rm{eV}$ and $R_{\rm{ED}}=5\times10^{-5}~\rm{cm}$. The oscillation channel is denoted in each subcaption. In all panels, the gray dashed and red solid curves are for the standard $3\nu$ scheme and the LED model, respectively.}
\end{figure}

Figures~\ref{fig:LEDprob-mubar} and \ref{fig:LEDprob-taubar} show the oscillation probabilities of $\bar{\nu}_\mu\to\bar{\nu}_\mu$ and $\bar{\nu}_\mu\to\bar{\nu}_\tau$, respectively. Correspondingly, Figures~\ref{fig:LEDprob-mu} and \ref{fig:LEDprob-tau} are for $\nu_\mu$ oscillation probabilities. In all the figures we assumed $m_1^D=0.01$~eV and $R_{\rm ED}=5\times10^{-5}$~cm, and the plots are for neutrinos passing the diameter of Earth, that is $\cos\theta_z=-1$. The gray dashed and red solid curves are for the standard $3\nu$ scheme and the LED model, respectively. The resonances discussed in Eq.~(\ref{eq:res}) can be seen in Figure~\ref{fig:LEDprob-mubar}. As we expected, the resonances exist just for $\bar{\nu}_\mu$. For $R_{\rm ED}=5\times10^{-5}$~cm, Eq.~(\ref{eq:Eres}) gives $(0.4,1.6,3.6,6.4,10)$~TeV for the first five resonance energies which match the position of dips in Figure~\ref{fig:LEDprob-mubar}. The decreasing depth of the dips for the higher KK modes is a consequence of the decreasing mixing angle between $\nu_{iL}^{(0)}$ and $\nu_{iL}^{(n)}$ ($\sin\vartheta_{n}\propto1/n$). The $\nu_\mu\to\nu_e$ and $\bar{\nu}_\mu\to\bar{\nu}_e$ oscillation probabilities are not shown since in both $3\nu$ scheme and the LED model the matter potential $V_e$ suppresses oscillation and the oscillation probability is zero for $E_\nu\gtrsim 0.1$~TeV. The nonzero oscillation probability $\bar{\nu}_\mu\to\bar{\nu}_\tau$ in Figure~\ref{fig:LEDprob-taubar}, showing as peaks at the resonance energies, are due to the $\bar{\nu}_\tau - \bar{\nu}_{s}^{(n)}$ mixings (we will discuss it in section~\ref{sec:3+n}, see also~\cite{Esmaili:2013cja,Choubey:2007ji}).

The oscillation probabilities for the trajectories passing the mantle ($\cos\theta_z\gtrsim-0.8$) are qualitatively similar to Figure~\ref{fig:LEDprob}, while the resonances are at $(0.64,2.56,5.76,10.24,16)$~TeV (for the same values of $m_1^D$ and $R_{\rm ED}$ as in Figure~\ref{fig:LEDprob}) and the dips are less profound due to the absence of the parametric resonance for these trajectories.    

In Figure~\ref{fig:LEDprob} the oscillation probabilities are shown for fixed values of $m_1^D$ and $R_{\rm ED}$. However, to confront the IceCube data with the expectation from LED model, we would scan all the parameter space of $(m_1^D,R_{\rm ED})$. We will report the result of this analysis in section~\ref{sec:icecube}. In the next section we elaborate on the interpretation of Figure~\ref{fig:LEDprob} in terms of the $(3+n)$ scenario.

\section{The equivalence between LED and $(3+n)$ models\label{sec:3+n}}

The KK modes in the LED model resemble a tower of sterile neutrinos from the brane point of view. There is a tower of sterile neutrinos for each flavor of the active neutrinos (or equivalently for each mass eigenstates of the active neutrinos), so an LED model with $n$ KK modes can be considered as a $(3+3n)$ model consisting of three active neutrinos and $3n$ sterile neutrinos. Translation of the LED model to a $(3+3n)$ model provides a better intuitive understanding of the results presented in the previous section, especially since there are already a rich literature on the oscillation pattern of the high energy atmospheric neutrinos in the $(3+1)$ model~\cite{Esmaili:2012nz,Esmaili:2013vza,Esmaili:2013cja,Razzaque:2011ab,Choubey:2007ji,Nunokawa:2003ep,Peres:2000ic}, which can be easily generalized to the $(3+3n)$ model. 

\begin{figure}[t!]
\centering
\subfloat[$P(\bar{\nu}_\mu\to\bar{\nu}_\mu)$]{
\includegraphics[width=0.5\textwidth]{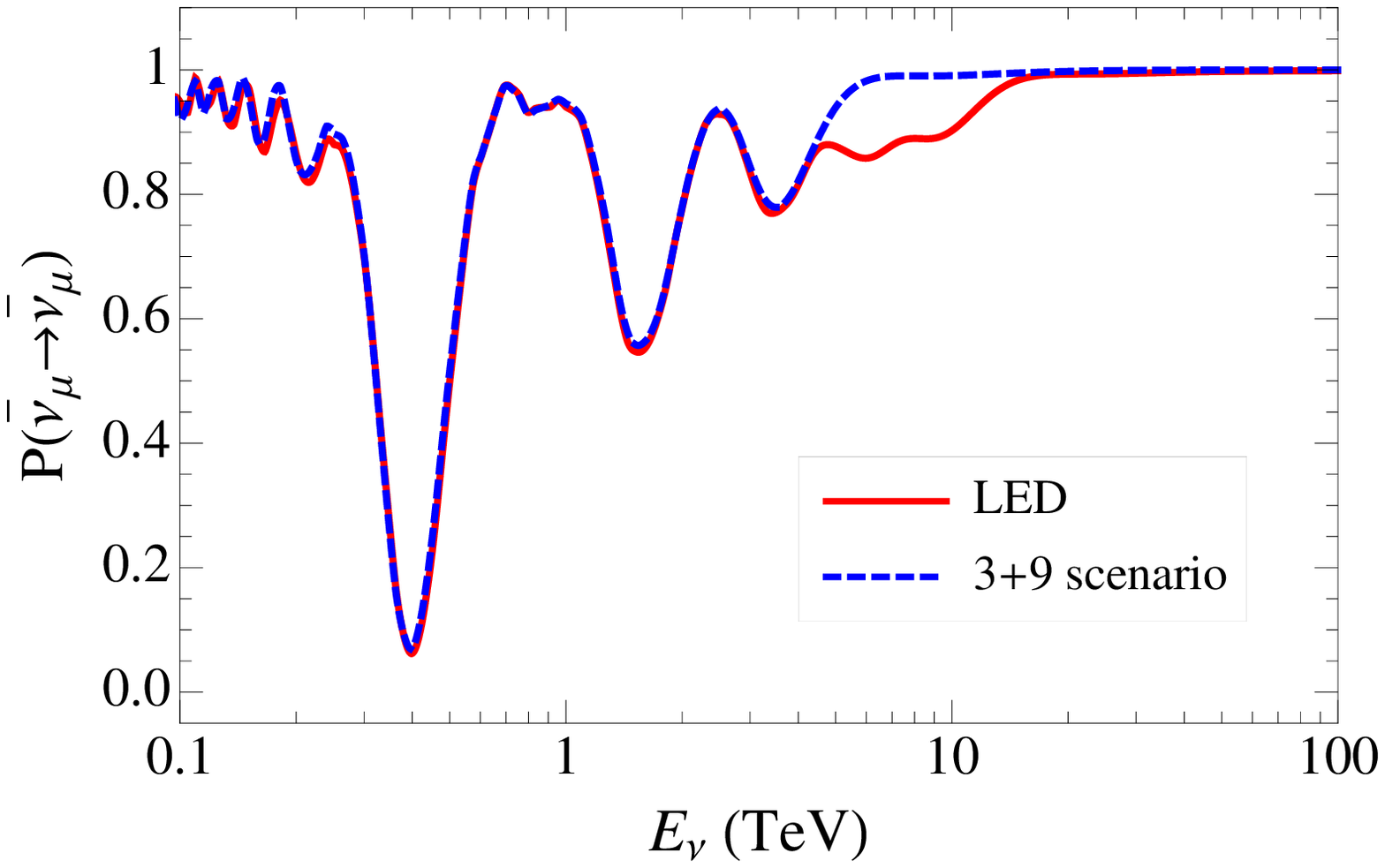}
\label{fig:compare-mu}
}
\subfloat[$P(\bar{\nu}_\mu\to\bar{\nu}_\tau)$]{
\includegraphics[width=0.5\textwidth]{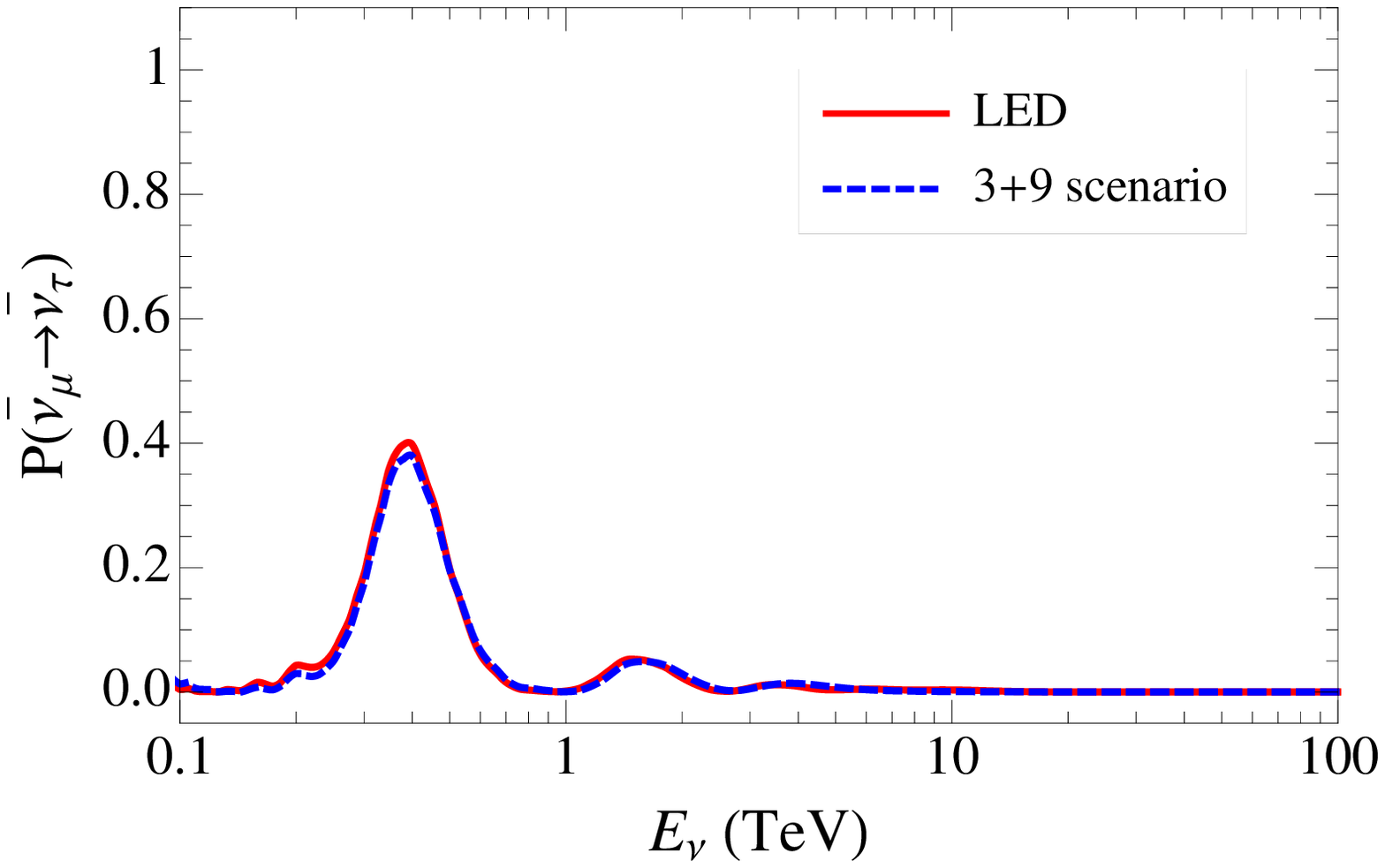}
\label{fig:compare-tau}
}
\caption{\label{fig:compare}Comparison of the oscillation probabilities calculated in the LED and the $(3+3n)$ models. For the LED model, we assume $5$ KK modes, $m_1^D=0.01$~eV and $R_{\rm ED}=5\times10^{-5}$~cm. For the $(3+3n)$ scenario we assume $n=3$ sterile neutrino states. Oscillation probabilities are for neutrinos passing the diameter of Earth ($\cos\theta_z=-1$). As can be seen up to the third KK mode the calculation in both models agree. By considering $(3+3n)$ scenario with larger $n$ this agreement extends to higher KK modes.}
\end{figure}

Let us briefly summarize the active-sterile mixing in the $(3+3n)$ model. The mixing matrix in this scenario is a $(3+3n)\times(3+3n)$ unitary matrix $W_{3+3n}$ which can be parametrized by $(3n+3)(3n+2)/2$ mixing angles\footnote{Among the $(3n+3)(3n+2)/2$ mixing angles, $3n(3n-1)/2$ angles quantify the sterile-sterile mixings of the $3n$ sterile states. So, since the sterile states do not enter the charged current interactions these angles are not relevant in the phenomenology of active neutrinos on the brane.} (we assume CP symmetry in lepton sector)
\begin{equation}\label{eq:wn}
W_{3+3n} = \prod_{j=2}^{3+3n} \left( \prod_{i=1}^{j-1} R_{ij} (\theta_{ij}) \right)~,
\end{equation}
where the ordered product is defined as $\prod_{i=1}^k A_i = A_k A_{k-1} \ldots A_1$, and $R_{ij} (\theta_{ij})$ is the rotation matrix in the $ij$ plane by the angle $\theta_{ij}$. The active flavor states $\nu_{\alpha L}$ are related to the mass eigenstates $\nu_j$ by
\begin{equation}\label{eq:3nflavor}
\nu_{\alpha L} = \sum_{j=1}^{3+3n} \left( W_{3+3n} \right)_{\alpha j} \nu_j~.
\end{equation}
By identifying $\nu^{\prime(q)}_{iL}\equiv\nu_{3q+i}$, comparison of Eq.~(\ref{eq:3nflavor}) and Eq.~(\ref{eq:nualpha}) enables us to derive the values of the elements of mixing matrix $W_{3+3n}$ in terms of the LED model parameters, which are $R_{\rm ED}$ and $m_1^D$. In order to elaborate on this equivalence between the LED model and the $(3+3n)$ model, in Figure~\ref{fig:compare} we compare the oscillation probabilities calculated in both models. In both panels of Figure~\ref{fig:compare} the red solid curve is for the LED model, the same as the one shown in Figure~\ref{fig:LEDprob}, with 5 KK modes. The dashed blue line correspond to $(3+3n)$ scenario with $n=3$. The dashed blue line is obtained by solving the following evolution equation     
\begin{equation}\label{eq:3+n}
i\frac{d\nu_{\alpha}}{dr} = \left[ \frac{1}{2E_\nu} W_{3+3n} \mathbf{M}^2 W_{3+3n}^\dagger + \mathbf{V}(r) \right]_{\alpha\beta} \nu_{\beta}~,
\end{equation}
where $\alpha,\beta=e,\mu,\tau,s_1,\ldots,s_{3n}$ (the $s_i$ is the $i^{\rm th}$ sterile flavor eigenstate). The elements of $W_{3+3n}$ obtained by comparing Eq.~(\ref{eq:3nflavor}) with Eq.~(\ref{eq:nualpha}); and $\mathbf{M}^2$ is a $(3+3n)\times (3+3n)$ diagonal matrix where the elements are mass-squared differences 
$$\mathbf{M}^2 = {\rm diag} \left( 0, \Delta m_{21}^2,\Delta m_{31}^2, \Delta m_{41}^2,\ldots, \Delta m_{3+3n,1}^2 \right)~,$$ 
where for the $q^{\rm{th}}$ KK mode ($q\geq1$ and we are assuming $m_1^D R_{\rm ED}\ll1$) 
$$\Delta m_{3+q,1}^2=\Delta m_{3+q+1,1}^2=\Delta m_{3+q+2,1}^2 = \frac{q^2}{R_{\rm ED}^2}~.$$
The potential matrix in Eq.~(\ref{eq:3+n}) is given by 
$$\mathbf{V}(r) = \sqrt{2}G_F{\rm diag} \left( n_e(r), 0,0, \frac{1}{2}n_n(r), \ldots, \frac{1}{2}n_n(r) \right)~.$$
Since we are assuming $n=3$ in $(3+3n)$ scenario, in the comparison of the LED model with $n=5$ KK modes the oscillation probabilities in both models should match up to the third KK mode and for higher KK modes deviations should appear.

Figures~\ref{fig:compare-mu} and \ref{fig:compare-tau} show the oscillation probabilities $P(\bar{\nu}_\mu\to\bar{\nu}_\mu)$ and $P(\bar{\nu}_\mu\to\bar{\nu}_\tau)$, respectively. As can be seen in panel (a), the probabilities match up to the third KK mode; the same is in panel (b), although since the peaks are very small the deviation in higher KK modes is not visible. In panel (a) clearly the deviation can be seen for the fourth and fifth KK mode resonances. Now, with the equivalence we are discussing in this section, it is easy to understand the peak in panel (b). It originates from the nonzero value of $(W_{3+3n})_{\tau j}$, that is the mixing between $\nu_\tau$ and the sterile states. This effective conversion of $\bar{\nu}_\mu\to\bar{\nu}_\tau$ when $(W_{3+3n})_{\tau j}\neq0$ has been already studied in the literature~\cite{Esmaili:2013cja,Choubey:2007ji}. In fact this effective conversion is the source of the sensitivity of cascade events in IceCube to $\theta_{3,3+3n}$ angles, which are poorly constrained by the current experiments (see the discussion in~\cite{Esmaili:2013cja}).

Although we discussed the equivalence between the LED model with $n$ KK modes and $(3+3n)$ scenario, this equivalence can be further simplified to $(3+n)$ scenario. As we mentioned in section~\ref{sec:formalism}, for $R_{\rm ED}\lesssim10^{-4}$~cm even the first KK mode states are much heavier than the active neutrino states and so effectively the three states $\nu_{1L}^{\prime(n)}$, $\nu_{2L}^{\prime(n)}$ and $\nu_{3L}^{\prime(n)}$ of the $n^{\rm th}$ KK mode are degenerate in mass. Thus, in principle it would be possible to redefine the states in each KK mode in such a way that, in two flavors approximation of active-sterile oscillation, just one of the new states mixes with the active neutrinos and the other two decouple. By this redefinition of states, the LED model with $n$ KK modes would be equivalent (at two flavors approximation) to the $(3+n)$ model, which has much fewer mixing parameters than the $(3+3n)$ model. In the following we elaborate on this equivalence and derive the corresponding effective mixing parameter values in the $(3+n)$ model.

In the phenomenology of high energy atmospheric neutrino oscillation in the presence of sterile neutrinos it is always possible to reduce the active-sterile mixing patterns to two-flavor systems of $\nu_e-\nu_{s_p}$, $\nu_\mu-\nu_{s_p}$ and $\nu_\tau-\nu_{s_p}$. In this approximation the oscillation of $\nu_e$, $\nu_\mu$ and $\nu_\tau$ flavors to $p^{\rm th}$ sterile state $\nu_{s_p}$ can be described by the effective mixing angles $\vartheta_{ep}$, $\vartheta_{\mu p}$ and $\vartheta_{\tau p}$ respectively. In the LED model the expansion of active flavor neutrino states in terms of the mass eigenstates in Eq.~(\ref{eq:nualpha}) can be written as 
\begin{equation}\label{eq:ledmixing}
\begin{pmatrix}
\nu_{eL} \\
\nu_{\mu L} \\
\nu_{\tau L}
\end{pmatrix}
= \sum_{n=0}^\infty U\mathcal{S}^{(n)} 
\begin{pmatrix}
\nu_{1L}^{\prime(n)} \\
\nu_{2L}^{\prime(n)} \\
\nu_{3L}^{\prime(n)}
\end{pmatrix},
\end{equation}
where $U$ is the PMNS matrix and $\mathcal{S}^{(n)}$ is a $3\times3$ diagonal matrix with the elements $\mathcal{S}^{(n)}={\rm diag}(S_1^{0n},S_2^{0n},S_3^{0n})$. For a fixed $n\geq1$ we can change the basis $(\nu_{1L}^{\prime(n)},\nu_{2L}^{\prime(n)},\nu_{3L}^{\prime(n)})$ to a new basis $(\tilde{\nu}_{1L}^{(n)},\tilde{\nu}_{2L}^{(n)},\tilde{\nu}_{3L}^{(n)})$ such that in this new basis just $\tilde{\nu}_{1L}^{(n)}$ contributes to $\nu_{\alpha L}$ state and the two states $\tilde{\nu}_{2L}^{(n)}$ and $\tilde{\nu}_{3L}^{(n)}$ decouple from the active neutrino $\nu_{\alpha L}$ and just contribute to sterile flavor states. Obviously, for $\nu_e$, $\nu_\mu$ and $\nu_\tau$ the new state $\tilde{\nu}_{1L}^{(n)}$ is proportional respectively to the first, second and third component of $U\mathcal{S}^{(n)}(\nu_{1L}^{\prime(n)},\nu_{2L}^{\prime(n)},\nu_{3L}^{\prime(n)})^T$ and the proportionality constant is given by the length of new basis. So, in the two-flavor system of $\nu_\alpha-\nu_{s_p}$ the effective mixing angle is given by (for $p\geq1$)
\begin{equation}\label{eq:t}
\sin \vartheta_{\alpha p} = \left[ \sum_{i=1}^3 \left| U_{\alpha i} S_{i}^{0p} \right|^2 \right]^{1/2}~.
\end{equation}

\begin{figure}[t!]
\centering
\subfloat[]{
\includegraphics[width=0.5\textwidth]{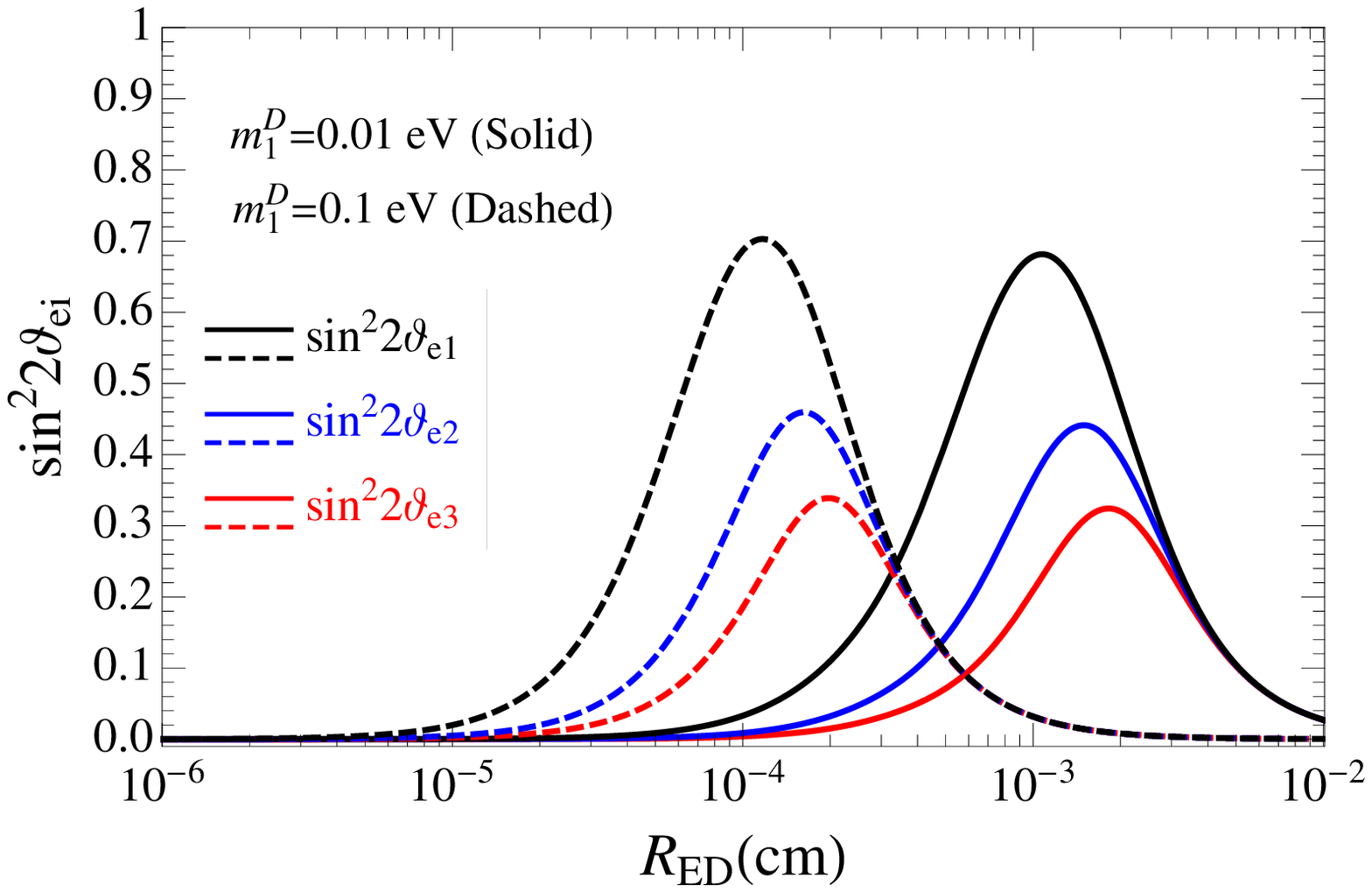}
\label{fig:theta-1n}
}
\subfloat[]{
\includegraphics[width=0.5\textwidth]{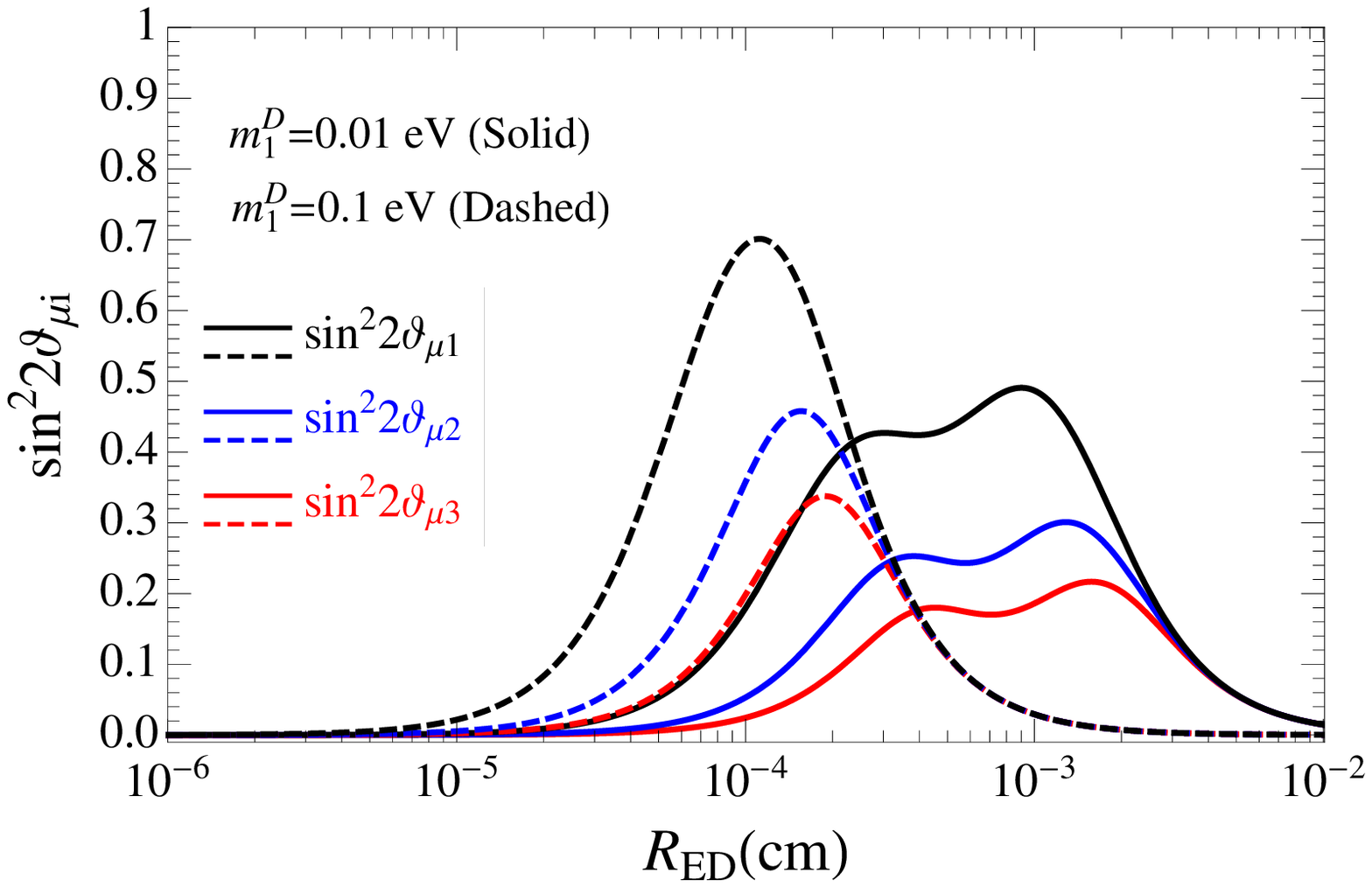}
\label{fig:theta-2n}
}
\quad
\subfloat[]{
\includegraphics[width=0.5\textwidth]{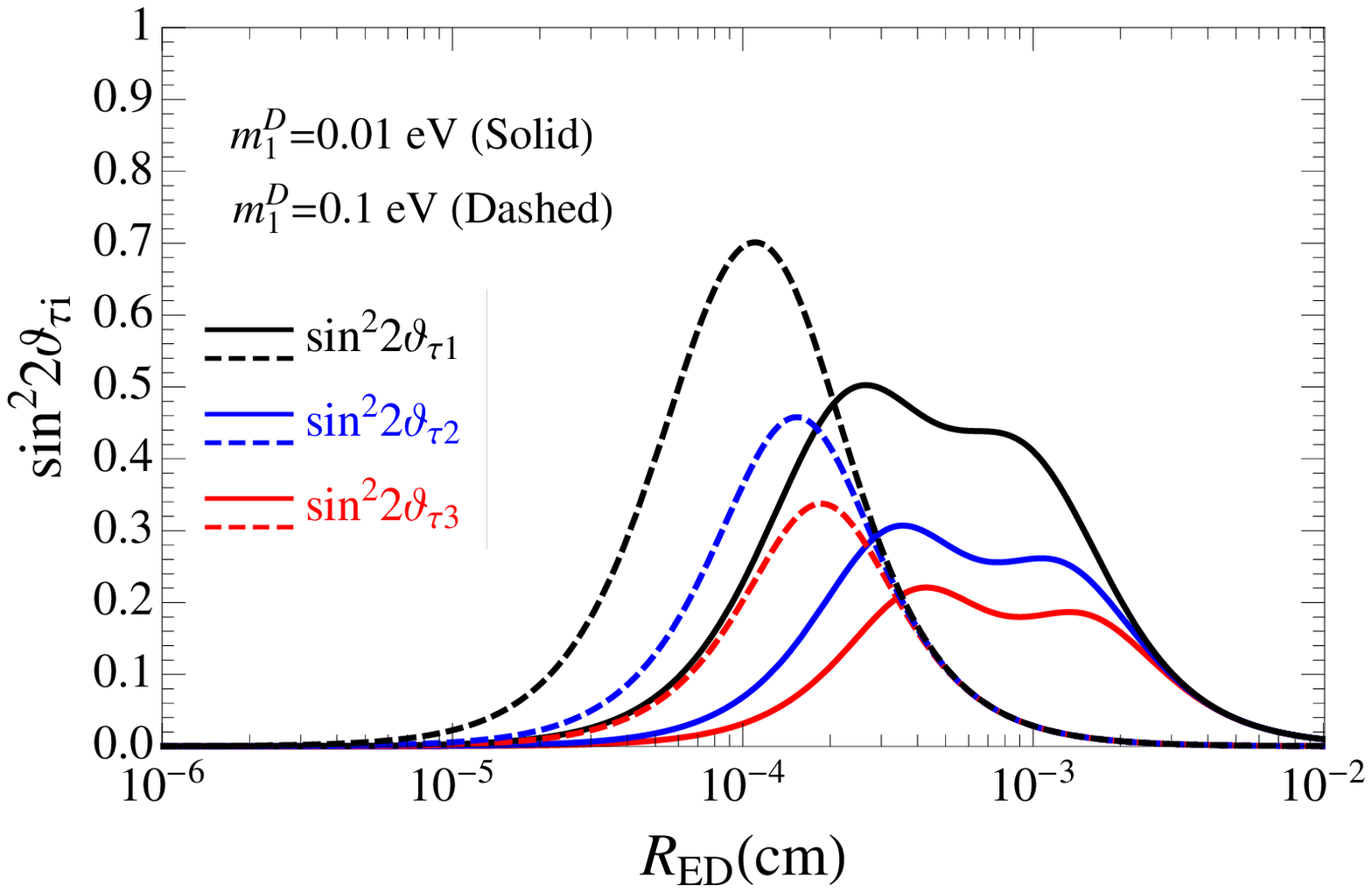}
\label{fig:theta-3n}
}
\caption{\label{fig:theta-in}The effective mixing angles $(\vartheta_{en},\vartheta_{\mu n},\vartheta_{\tau n})$ in the $(3+3)$ scenario which is equivalent to the LED model with 3 KK modes. In all the plots the solid and dashed curves correspond respectively to $m_1^D=0.01$~eV and $0.1$~eV.}
\end{figure}

Figure~\ref{fig:theta-in} shows the corresponding effective active-sterile mixing angles of the $(3+3)$ scenario equivalent to the LED model with 3 KK modes ($n=3$), as function of $R_{\rm ED}$ for $m_1^D=0.1$~eV (dashed curves) and $0.01$~eV (solid curves). The peak-shape behavior of the curves in all the panels originate from the behavior of $S_i^{0n}$. It can be shown from Eq.~(\ref{eq:s0n}) that the maxima of $S_i^{0n}$ occur at $m_i^D R_{\rm ED}\simeq \sqrt{n/\pi}$ (we will call it maximum condition hereafter). In fact in each curve of Figure~\ref{fig:theta-in} there are three peaks at values of $R_{\rm ED}$ derived from\footnote{Notice that there are few percent uncertainties in this computation since we are approximating $\lambda_i^{(n)}\simeq n$; while more accurately $\lambda_i^{(n)}$ is a number between $n$ and $n+1/2$.} $m_i^D R_{\rm ED}\simeq \sqrt{n/\pi}$ for $i=1,2,3$; and the relative heights of these peaks are controlled by the relative size of $U_{\alpha i}$ (where $\alpha=e$, $\mu$ and $\tau$, respectively for $\vartheta_{en}$, $\vartheta_{\mu n}$ and $\vartheta_{\tau n}$). However, as far as $m_1^D\gtrsim\sqrt{\Delta m_{\rm atm}^2}$ these three peaks coincide and effectively one peak can be recognized. This coincidence of the peaks can be seen for the case $m_1^D=0.1$~eV depicted by the black dashed curves in Figure~\ref{fig:theta-in}. For $m_1^D=0.1$~eV and $n=1$, from maximum condition we obtain $R_{\rm ED}\simeq1.1\times10^{-4}$~cm which agrees with the peak's positions of black dashed curves in Figure~\ref{fig:theta-in}. For higher KK modes the peak position slightly moves to larger $R_{\rm ED}$. The separation of peaks is visible for smaller values of $m_1^D$. Let us consider the case $m_1^D=0.01$~eV depicted by solid curves in Figure~\ref{fig:theta-in}. In this case $m_2^D\simeq m_1^D$ while $m_3^D\simeq\sqrt{\Delta m_{\rm atm}^2}=0.05$~eV. Thus, for $n=1$, the maximum condition leads to two peaks at $R_{\rm ED}\simeq2.3\times10^{-4}$~cm and $1.1\times10^{-3}$~cm which clearly can be identified in the black solid curves of Figures~\ref{fig:theta-2n} and \ref{fig:theta-3n}. The first peak (which is due to $m_3^D$) is not visible in the black solid curve of Figures~\ref{fig:theta-1n} since it is suppressed by the small value of $U_{e3}$. For higher $n$, again the peaks slightly move to larger $R_{\rm ED}$ (compare different colors of solid curves in each panel). For $m_1^D\lesssim\sqrt{\Delta m_{\rm sol}^2}\simeq9\times10^{-3}$~eV a third peak in large values of $R_{\rm ED}$ will develop. However, notice that for $m_1^D\to0$ the position of peaks originating from $m_2^D=\sqrt{\Delta m_{\rm sol}^2}$ and $m_3^D=\sqrt{\Delta m_{\rm atm}^2}$ do not change, which means that always there are two peaks at $R_{\rm ED}=2.3\times10^{-4}$~cm and $1.3\times10^{-3}$~cm for both $\vartheta_{\mu n}$ and $\vartheta_{\tau n}$. Thus, we can immediately conclude that for $m_1^D\lesssim10^{-2}$~eV the sensitivity of IceCube to the LED model is independent of the value of $m_1^D$. By inspecting the black solid curve in Figure~\ref{fig:theta-2n}, it can be seen that $\sin^22\vartheta_{\mu 1}\simeq0.1$ for $R_{\rm ED}\simeq5\times10^{-5}$~cm and so IceCube would be able to constrain $R_{\rm ED}$ at this level for $m_1^D\lesssim10^{-2}$~eV.  

Let us discuss the case of $m_1^D\gtrsim0.1$~eV. In this case, as can be seen also from the dashed curves in Figure~\ref{fig:theta-in}, all the three peaks coincide (since $m_1^D\simeq m_2^D\simeq m_3^D$) at $R_{\rm ED}\simeq \sqrt{n/\pi}/m_1^D$. This means that by increasing $m_1^D$, IceCube will be sensitive to smaller values of $R_{\rm ED}$ such that the sensitivity contour in the log-log plot of $(R_{\rm ED},m_1^D)$ plane will be a straight line with the slope $-1$. The intercept of this line can be estimated from Figure~\ref{fig:theta-in}. From the black dashed curve in Figure~\ref{fig:theta-2n}, it can be seen that $\sin^22\vartheta_{\mu 1}\simeq0.1$ for $R_{\rm ED}\simeq2\times10^{-5}$~cm. From this we conclude that IceCube would be able to constrain LED radius down to $R_{\rm ED} \simeq 2\times10^{-6}({\rm eV}/m_1^D)$~cm for $m_1^D\gtrsim0.1$~eV. We should mention that large values of $m_1^D$ have severe conflicts with the bounds on neutrino mass from cosmological considerations such that $m_1^D\gtrsim1$~eV can be ruled out robustly~\cite{Ade:2013zuv}.

\section{Constraining the LED model with the IceCube data\label{sec:icecube}}

In section~\ref{sec:osc} we calculated the flavor oscillation probabilities of high energy atmospheric neutrinos in the LED model. In this section we analyze the collected atmospheric data in IceCube to search for the signatures of LED model in the zenith distribution of events. Although, as we have shown in section~\ref{sec:3+n}, the oscillation probabilities can be calculated in the equivalent $(3+3n)$ or $(3+n)$ scenarios, for the analysis of this section calculations have been done in the original LED model assuming 5 KK modes. However, for the interpretation of results obtained in this section, we extensively use the terminology of $(3+n)$ scenario, that is the effective mixing angles in  Eq.~(\ref{eq:t}).   

We analyze two sets of the IceCube data, IC-40~\cite{Abbasi:2010ie} and IC-79~\cite{Aartsen:2013jza}, consisting of the muon-track events induced by atmospheric neutrinos respectively in the energy range $(0.1-400)$~TeV and $(0.1-10)$~TeV. These data sets provide the zenith distribution of events and so in our analysis we would consider just the integrated number of events over the energy. We will discuss later the improvements that can be achieved by adding the energy information of events. The number of muon-track events in the $i^{\rm th}$ bin of zenith angle $\Delta_i\cos\theta_z$ can be calculated by
\begin{equation}\label{eq:nmu}
N_i= T\Delta\Omega \sum_{\alpha=e,\mu} \left\{ \int dE_\nu \int_{\Delta_i} d\cos\theta_z A^{\nu}_{\rm{eff}}(E_\nu,\cos\theta_z) \Phi_{\nu_{\alpha}}(E_\nu,\cos\theta_z) P(\nu_{\alpha}\to \nu_{\mu})\right\} +(\nu\to\bar{\nu})~,
\end{equation}
where $T$ is the data-taking period, 359 and 319 days respectively for IC-40 and IC-79; $\Delta\Omega=2\pi$ is the azimuthal acceptance of IceCube detector, $\Phi_{\nu_\alpha}$ is the atmospheric $\nu_\alpha$ flux taken from~\cite{Honda:2006qj} and $A_{\rm eff}^\nu$ is the neutrino effective area which for IC-40 and IC-79 we take respectively from~\cite{Esmaili:2012nz} and \cite{Esmaili:2013fva}. Finally, the $P(\nu_\alpha\to\nu_\mu)$ in Eq.~(\ref{eq:nmu}) is the neutrino oscillation probability which is discussed in section~\ref{sec:osc}. Although the $\nu_e$ and $\bar{\nu}_e$ atmospheric fluxes at high energies are quite small, we consider them for the sake of completeness.   

To confront the LED model with the IceCube data and probing the LED parameters, we define the following $\chi^2$ function:
\begin{eqnarray}\label{chi2}
\chi^2 \left( m_1^D,R_{\rm{ED}};\alpha,\beta\right) & = & \sum^{10}_{i=1} \frac{\left\{N^{\rm data}_i - \alpha\left[ 1+\beta\left(0.5+(\cos\theta_z)_i\right)\right] N_i(m_1^D,R_{\rm ED})\right\}^2}{\sigma^2_{i,\rm{stat}}+\sigma^2_{i,\rm{sys}}}\nonumber\\
& + & \frac{(1-\alpha)^2}{\sigma^2_\alpha}+\frac{\beta^2}{\sigma^2_\beta},
\end{eqnarray}
where $N_i^{\rm data}$ is the observed number of events in the $i^{\rm th}$ bin of the zenith angle $\Delta_i\cos\theta_z$. For both IC-40 and IC-79 we take 10 equal bins of zenith angle and so the up-going muon-track events are divided to zenith bins with width $\Delta_i\cos\theta_z=0.1$. The $N_i(m_1^D,R_{\rm ED})$ is the expected number of events in the $i^{\rm th}$ bin, given by Eq.~(\ref{eq:nmu}), in the LED model with parameters $m_1^D$ and $R_{\rm ED}$. The parameters $\alpha$ and $\beta$ take into account respectively the correlated systematic uncertainties of the normalization and the tilt of atmospheric neutrino flux, with $\sigma_{\alpha}=0.24$ and $\sigma_{\beta}=0.04$~\cite{Honda:2006qj}. The $\sigma_{i,\rm{stat}}=\sqrt{N_i^{\rm data}}$ is the statistical error and $\sigma_{i,\rm{sys}}=fN_i$ is the uncorrelated systematic uncertainty quantified by the parameter $f$, where $f=7\%$ for IC-40 and $f=2\%$ for IC-79. Marginalization of $\chi^2$ function with respect to $\alpha$ and $\beta$ gives the constraint on the LED parameters. The LED model do not improve the fit to the data as can be seen by comparing the values of $\chi^2$ at best-fit points in the standard $3\nu$ scheme and the LED model reported in Table~\ref{tab:1}. Thus, the data of IceCube can be used to constrain the LED parameters. 

\begin{table}
\caption{\label{tab:1}Comparing the goodness of fit between the $3\nu$ scheme and the LED model for IC-40 and IC-79 data sets.}
\vspace{0.5cm}
\begin{tabular}{|c|c|c|}
\hline
data set & $\chi^2_{3\nu,\rm{min}}$ & $\chi^2_{{\rm LED},\rm{min}}$\\
\hline
IC-40 & 10.1 & 9.7 \\
\hline
IC-79 & 8.9 & 9.0 \\
\hline
\end{tabular}
\vspace{0.5cm}
\end{table}

\begin{figure}[t!]
\centering
\includegraphics[width=0.7\textwidth]{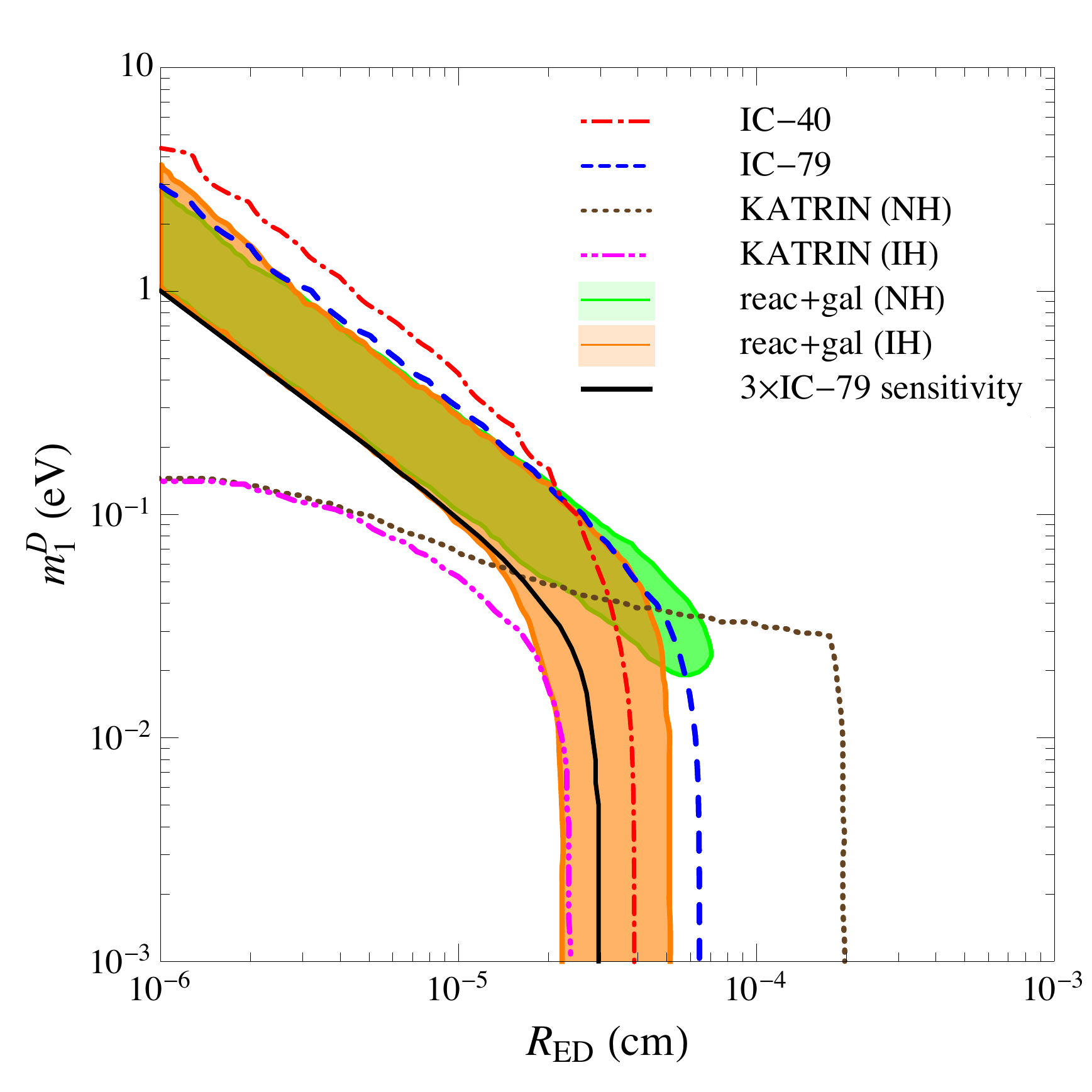}
\caption{\label{fig:limit}The allowed regions for LED model in the plane $(R_{\rm ED},m_1^D)$. The red dashed and blue dot-dashed curves are obtained from the analyses of IC-40 and IC-79 data sets, respectively, at $2\sigma$ C.L.. The green and orange shaded regions are the preferred regions by reactor and gallium anomalies at $2\sigma$ C.L., respectively for NH and IH, taken from~\cite{Machado:2011kt}. The brown dotted and violet dashed curves show the sensitivity of KATRIN, at $90\%$ C.L., respectively for NH and IH, taken from~\cite{BastoGonzalez:2012me}. The black solid curve shows the sensitivity of IceCube to LED model at $99\%$ C.L., assuming 3 times larger statistics than IC-79 and taking into account the energy information of events.}
\end{figure}

Figure~\ref{fig:limit} shows the allowed region in the plane $(R_{\rm ED},m_1^D)$ from the analysis of IceCube data. The red dot-dashed and blue dashed curves show the $2\sigma$ contours obtained from IC-40 and IC-79 data sets respectively. As we discussed in section~\ref{sec:3+n}, these contours consist of two parts: a vertical part for $m_1^D \lesssim 0.1$~eV and a straight line with slope -1 for $m_1^D\gtrsim0.1$~eV. In section~\ref{sec:3+n} we estimated also the position of these parts, that is the intercepts of these lines: $R_{\rm ED} \simeq5\times10^{-5}$~cm for $m_1^D \lesssim 0.1$~eV and $R_{\rm ED}\simeq2\times10^{-6}$~cm for $m_1^D \simeq 1$~eV which are in agreement with Figure~\ref{fig:limit}.  

In Figure~\ref{fig:limit} the green and orange shaded regions show the $2\sigma$ level preferred values of $m_1^D$ and $R_{\rm ED}$ from reactor and gallium anomalies, respectively for NH and IH, taken from~\cite{Machado:2011kt}. The brown dotted and purple double-dot-dashed curves show the sensitivity of KATRIN experiment to the LED parameters at $90\%$ C.L., respectively for NH and IH, taken from~\cite{BastoGonzalez:2012me}. Finally, the black solid curve shows the sensitivity of IceCube at $99\%$ C.L. by considering the energy information of events and assuming 3 times of IC-79 data, which is available now. In the following we discuss each of the components in Figure~\ref{fig:limit} and their implications. In fact the equivalence of the LED and the $(3+n)$ models, constructed in section~\ref{sec:3+n}, helps us to easily interpret Figure~\ref{fig:limit}. 

Very concisely, the reactor~\cite{Mueller:2011nm} and gallium~\cite{Giunti:2010zu} anomalies are respectively the deficits in the number of events observed in the short baseline reactor and calibration of the solar neutrino experiments, which point to $P(\nu_e(\bar{\nu_e})\to\nu_e(\bar{\nu_e}))\neq1$ over short distances that obviously cannot be accommodated in the standard $3\nu$ scheme. These deficits can be interpreted in the $(3+n)$ scenario by the $\nu_e-\nu_s$ mixing that leads to the oscillation of $\nu_e$ and $\bar{\nu}_e$ to the sterile neutrino states which escape from detection in the detectors~\cite{Kopp:2013vaa}. Thus, reactor and gallium anomalies require $\vartheta_{en}\neq0$ or in the simplest $(3+1)$ scenario $\vartheta_{e1}\equiv\theta_{14}\neq0$. In a generic $(3+n)$ scenario the mixing angles $\vartheta_{en}$, $\vartheta_{\mu n}$ and $\vartheta_{\tau n}$ are independent parameters that can take any value. On the other hand, the IceCube muon-track data is not sensitive to $\vartheta_{en}$ angles. Also, as it is shown in~\cite{Esmaili:2013cja}, IceCube cascade data is not sensitive to the values of $\vartheta_{en}$ preferred by reactor and gallium anomalies. Thus, in a generic $(3+n)$ scenario for the interpretation of these anomalies, IceCube cannot provide an independent check. However, this is not the case in the LED model. For the LED model, all the angles in the equivalent $(3+n)$ scenario are inter-related and non-vanishing $\vartheta_{en}$ lead to non-vanishing $\vartheta_{\mu n}$ and $\vartheta_{\tau n}$. Thus, since the IceCube muon-track data can probe $\vartheta_{\mu n}$ and $\vartheta_{\tau n}$, it is possible to probe the LED interpretation of reactor and gallium anomalies which have been proposed in~\cite{Machado:2011kt}. As can be seen from Figure~\ref{fig:limit}, the IC-40 and IC-79 data can exclude a part of the preferred region by these anomalies. 

It is possible to probe the green and orange shaded regions in Figure~\ref{fig:limit} by considering the energy information of IceCube data. Since the energy information of IceCube data is not publicly available we estimate the sensitivity of IceCube assuming a data set 3 times the IC-79 data set (which already are collected). The sensitivity of IceCube to the sterile neutrinos after taking into account the energy information has been calculated in~\cite{Esmaili:2013vza}. From the Figure~10 of~\cite{Esmaili:2013vza} it can be seen that, by considering the energy information, IceCube can probe the sterile neutrino mixing $\sin^22\vartheta_{\mu1}\simeq0.02$ for $\Delta m_{41}^2\lesssim1~{\rm eV}^2$. Using the equivalence constructed in section~\ref{sec:3+n} this sensitivity can be translated to the sensitivity of IceCube to the LED model. From the mixing angles plotted in Figure~\ref{fig:theta-in}, we can check that $\sin^22\vartheta_{\mu1}\simeq0.02$ at $R_{\rm ED}\simeq3\times10^{-5}$~cm for $m_1^D\lesssim0.1$~eV; and at $R_{\rm ED} \simeq10^{-6}({\rm eV}/m_1^D)$~cm for $m_1^D\gtrsim0.1$~eV, which are in agreement with the black solid curve in Figure~\ref{fig:limit}. As can be seen, although the current data exclude only a small part of the region allowed by the reactor and Gallium anomalies, considering the energy information of atmospheric neutrino data can almost exclude all the favored regions (or to confirm the interpretation of these anomalies in terms of LED model). Performing such an analysis (i.e., taking into account the energy binning) requires detailed information of IceCube detector which is not available now. But, however, with the already collected data IceCube collaboration can perform this analysis.

The other way of probing the regions preferred by reactor and gallium anomalies is the KATRIN experiment (the brown dotted and purple double-dot-dashed curves in Figure~\ref{fig:limit}). As can be seen, for both NH and IH cases, the KATRIN can completely exclude the green and orange shaded regions. 

\section{Conclusions\label{sec:conc}}

An added bonus of the LED model is the explanation of small neutrino masses which can be achieved by introducing singlet fermions living in the bulk of extra dimensions. From the brane point of view these fermions constitute towers of sterile neutrinos with increasing masses (the so-called KK modes) that mix with the active neutrinos and so can affect the phenomenology of neutrino flavor oscillations. In fact, this picture can be favored due to the recent observed anomalies in the short baseline oscillation experiments which hint on the presence of one (or more) sterile neutrino state(s). On the other hand, the existence of these sterile neutrinos can significantly change the oscillation pattern of high energy atmospheric neutrinos observed by the IceCube experiment. In this paper we studied these effects and developed a framework to interpret them. 

The mixing of the KK modes of the bulk fermions with the active neutrinos lead to resonant conversion of $\bar{\nu}_\mu$ to the undetectable sterile neutrinos at high energies. The resonance originates from the matter effects (constant density MSW resonance) during the propagation of atmospheric neutrinos through the Earth and would lead to distortions in the zenith and energy distributions of muon-track events at the IceCube detector. IceCube has already published two sets of the atmospheric neutrino data (IC-40 and IC-79) and in this paper we analyzed them in the search of features predicted by the LED model.   

We obtained the limits on the LED parameters (especially the radius of extra dimension $R_{\rm ED}$) by analyzing the zenith distributions of IC-40 and IC-79 data. For $m_1^D\lesssim0.1$~eV the upper limit $R_{\rm ED}\leq4\times10^{-5}$~cm (at $2\sigma$ level) have been set by the IceCube data and is independent of the value of $m_1^D$. For $m_1^D\gtrsim0.1$~eV the limit depends on the value of $m_1^D$ and is stronger: $R_{\rm ED}\lesssim3\times10^{-6}({\rm eV}/m_1^D)$~cm. These bounds can exclude some parts of the parameter space preferred by the reactor and gallium anomalies.  

We have also discussed the prospect of improving the bounds by taking into account the energy distribution of muon-track events in the IceCube. We have shown that with a sample of data three times larger than the IC-79 data set (which is already collected by the IceCube detector from its completion at December/2010 till now) it would be possible to exclude the $2\sigma$ preferred region by the reactor and gallium anomalies.

As a tool for interpreting the obtained results in this paper, we developed an equivalence between the LED model and the phenomenological $(3+n)$ scenarios which have been studied extensively in the literature. This equivalence provides a clear and intuitive picture of the oscillation pattern of atmospheric neutrinos in the LED model and have been used in this paper to explain the features obtained by the numerical calculations.

\begin{acknowledgments}
For A.~E. this research was supported by the Munich Institute for Astro- and Particle Physics (MIAPP) of the DFG cluster of excellence ``Origin and Structure of the Universe". A.~E. thanks Nordita for hospitality and support during the program ``What is the Dark MAtter?". O.~L.~G.~P. thanks the ICTP and the financial support from the funding grant 2012/16389-1, S\~ao Paulo Research Foundation (FAPESP). A.~E. thanks the financial support from the grant Jovem Pesquisador 1155/13 by FAEPEX/UNICAMP and grant 1280477 from PNPD/CAPES. We thank the authors of Ref.~\cite{Machado:2011kt} for providing us the numerical tables of the allowed regions of their work.
\end{acknowledgments}


{}
  
\end{document}